\documentclass[twocolumn]{jpsj3}
\usepackage{txfonts}
\usepackage{graphicx, amsmath, amssymb,bm, mathrsfs, color, comment}
\graphicspath{{./figures/}}

\title{
Anisotropic $B$-$T$ Phase Diagram of Non-Kramers System PrRh$_2$Zn$_{20}$
}
\author{
Taichi~Yoshida$^1$\thanks{yoshida.t.ba@m.titech.ac.jp}, Yo~Machida$^1$, Koichi~Izawa$^1$, Yuki~Shimada$^2$, Naohiro~Nagasawa$^2$, Takahiro~Onimaru$^2$, Toshiro~Takabatake$^2$, Adrien~Gourgout$^3$, Alexandre~Pourret$^3$, Georg~Knebel$^3$, and Jean-Pascal~Brison$^3$
}
\inst{
$^1$Department of Physics, Tokyo Institute of Technology, Meguro, Tokyo 152-8551, Japan\\
$^2$Department of Quantum Matter, Graduate School of Advanced Sciences of Matter, Hiroshima University, Higashi-Hiroshima, Hiroshima 739-8530, Japan\\
$^3$Universit\'e Grenoble Alpes, CEA, INAC-PHELIQS, F-38000 Grenoble, France
}

\abst{
We have investigated the low temperature quadrupolar phenomena of the non-Kramers system PrRh$_2$Zn$_{20}$ under magnetic fields in the [100] and [110] directions. Our experiments reveal the $B$-$T$ phase diagram of PrRh$_2$Zn$_{20}$ involving four electronic states regardless of the field direction, namely, a non-Fermi liquid (NFL) state, an antiferro-quadrupolar (AFQ) ordered state, a novel heavy-fermion (HF) state, and a field-induced singlet (FIS) state. In the wide range of the NFL state, the resistivity can be well scaled by a characteristic temperature, suggesting the realization of the quadrupole Kondo effect. The HF state exhibits a Fermi liquid behavior with a large $A$ coefficient of the $T^2$ term in the resistivity, suggesting the formation of nontrivial heavy quasi-particles. The FIS state results from the considerable splitting of a non-Kramers doublet by a magnetic field. The phase diagram shows a large anisotropy with respect to the field direction. It is found that the anisotropy of the phase diagram can be explained in terms of that of the energy splitting of the non-Kramers doublet by a magnetic field. This indicates that the low temperature properties of PrRh$_2$Zn$_{20}$ are governed by the non-Kramers doublet, namely, quadrupole degrees of freedom. Since a similar phase diagram has been obtained for the related compound PrIr$_2$Zn$_{20}$, it is expected that the $B$-$T$ phase diagram constructed in this work is universal throughout non-Kramers systems.
}

\begin{document}
\maketitle

\section{Introduction}
\label{sec:intro}
In strong correlated $f$-electron systems, the itinerant-localized dual character of $f$-electrons, which is derived from the atomic $f$-orbitals via the hybridization between conduction and $f$-electrons ($c$-$f$ hybridization), gives rise to a rich variety of exotic phenomena at very low temperatures, for instance, magnetic order, heavy-fermion behavior, and quantum criticality. The vast majority of such phenomena are governed by the spin degrees of freedom and have been explained in terms of the interplay between the Kondo effect and Ruderman--Kittel--Kasuya--Yosida (RKKY) interaction. In particular, many attractive phenomena such as metamagnetic transition, unconventional superconductivity, and non-Fermi liquid behavior are realized on the verge of magnetic order. In other words, these phenomena have been systematically understood on the basis of the Doniach phase diagram in terms of the spin degrees of freedom. On the other hand, the nature of electronic states characteristic of multipole instead of spin is hardly unveiled at present, although the multipole is one of the fundamental degrees of freedom of $f$-electrons as well as spins. Only the long-range order of multipole realized in $f$-electron systems with weak $c$-$f$ hybridization has been well discussed. In fact, the quadrupole and octupole orders have been experimentally observed in, e.g., Ce$_3$Pd$_{20}$Ge$_6$~\cite{Kitagawa1998PRB}, Ce$_x$La$_{1-x}$B$_6$~\cite{Mannix2005PRL, Matsumura2009PRL}, and NpO$_2$~\cite{Paixao2002PRL}, and the hexadecapole order in PrRu$_4$P$_{12}$ has been proposed~\cite{Takimoto2006JPSJ}. In contrast, most of the low temperature properties inherent to multipoles with strong hybridization have not been clarified so far, even though the itinerant character of multipoles caused by the $c$-$f$ hybridization has been recognized to be a crucial factor to reveal intriguing physics such as the hidden order in URu$_2$Si$_2$~\cite{Chandra2013N, Ikeda2012NP}. Therefore, our interest has been directed to understanding the fundamental nature of electronic states in a non-Kramers system with strong hybridization.

One may expect that we can readily understand the properties of a non-Kramers system on the analogy of the Doniach picture. In fact, however, it is not utterly straightforward because the scattering process of a multipole is quite different from that of a spin. For example, an electric quadrupole moment, which is a rank-2 multipole, scatters with conduction electrons via two equivalent channels arising from the time-reversal symmetry. Hence, in contrast to a single-channel Kondo effect, an anomalous fixed point occurs at a finite coupling and the quadrupole moment is imperfectly screened (overscreening)~\cite{Nozieres1980JPP, Cox1987PRL, Cox1998AP}. Since the overscreening strongly prevents the formation of quasi-particles, the electronic state arising from such a scattering process with a quadrupole moment can no longer be described by Landau's Fermi liquid (FL) theory. In other words, a non-Kramers system is expected to show a non-Fermi liquid (NFL) state with anomalous temperature dependences of physical quantities such as $C/T\propto -\log T$ and $\Delta \rho \propto \sqrt{T}$~\cite{Affleck1993PRB}. According to recent theoretical studies based on the two-channel Anderson lattice model~\cite{Tsuruta2015JPSJ}, the temperature dependences of the NFL behaviors are expressed as
\begin{gather}
\Delta \rho (T) \equiv \rho (T)-\rho_0=\frac{a_1}{1+a_2 (T_0/T)},
\label{eq:tsuruta}\\
C(T) = b_1\left( 1-b_2\sqrt{T/T_0} \right),
\label{eq:tsuruta_C}
\end{gather}
where $a_i$ and $b_i$ ($i=1,2$) are constants, and $T_0$ represents the characteristic temperature of the two-channel Kondo lattice. Note that this NFL state necessarily has a residual entropy $\log \!\! \sqrt{2}$ arising from the equivalency of two scattering channels even at $T=0$. Hence, a novel electronic ground state to release the residual entropy at the low temperature limit is naturally expected instead of the NFL state. These features of the non-Kramers system are essentially different from the expectations from the Doniach phase diagram. Therefore, we would not be able to understand the nature of the non-Kramers system on the mere analogy of the Doniach picture for the Kramers system. 

The nature of the non-Kramers system is difficult to clarify mainly because of the lack of suitable compounds. The well-known compounds for candidate of the quadrupole Kondo effect include UBe$_{13}$~\cite{McElfresh1993PRB}, Y$_{1-x}$U$_{x}$Pd$_3$~\cite{Seaman1992JAC}, and U$_x$Th$_{1-x}$Ru$_2$Si$_2$~\cite{Amitsuka1994JPSJ}. However, it is unclear whether they can be regarded as a non-Kramers system because an uncertainty of the crystalline electric field (CEF) ground state is involved. For the other candidates PrMg$_3$~\cite{Tanida2006JPSJ} and PrAg$_2$In~\cite{Yatskar1996PRL}, the precise argument on the nature of the non-Kramers system may be hindered by the local breaking of the cubic symmetry due to the random site exchange inherent to Heusler-type compounds. Recently, triggered by the discovery of the new non-Kramers systems Pr$Tr_2X_{20}$ ($Tr$: transition metal, $X=$ Zn, Al), which are the so-called Pr 1-2-20 systems, investigations of the electronic properties of non-Kramers systems have markedly progressed for several years. As will be described in the next section, much effort has been devoted to studying the low temperature electronic states characteristic of the quadrupole degrees of freedom for mainly PrIr$_2$Zn$_{20}$, PrV$_2$Al$_{20}$, and PrTi$_2$Al$_{20}$. As a result, new exotic physics such as an anomalous NFL behavior, an unconventional FL state with a large electron mass, and the coexistence of quadrupole ordering and superconductivity have been discovered~\cite{Onimaru2011PRL, Onimaru2016PRB, Yoshida2015PP, Tsujimoto2014PRL, Sakai2011JPSJ, Matsubayashi2012PRL}. It can be said that these findings have opened the door toward the successive clarification of non-Kramers systems.

\section{Basic Properties of PrRh$_2$Zn$_{20}$ and Related Compounds}
\label{sec:compound}
Pr$Tr_2X_{20}$ are CeCr$_2$Al$_{20}$-type cubic compounds with the space group $Fd\bar{3}m$. Since the Pr$^{+3}$ ion is surrounded by 16 $X$ ions forming a Frank--Kasper cage, the strong $c$-$f$ hybridization is expected. In general, if a rare-earth ion with $f^2$ configuration such as Pr$^{3+}$ and U$^{4+}$ is subjected to a cubic CEF, a nonmagnetic (non-Kramers doublet) ground state with quadrupole degrees of freedom is sometimes realized. In practice, the Pr 1-2-20 systems have been confirmed to possess the non-Kramers $\Gamma_3$ doublet at the CEF ground state~\cite{Onimaru2011PRL, Onimaru2012PRB, Sakai2011JPSJ, Sato2012PRB, Iwasa2013JPSJ}. Since the CEF first excited energies are estimated to be about 30~K (Ir, Rh), 40~K (V), and 60~K (Ti), we would consider that the non-Kramers doublet is well isolated and the quadrupole degrees of freedom is predominantly active at very low temperatures. The non-Kramers doublet is known to carry no magnetic dipole moment but two components of quadrupole moment $\left( O_2^0, O_2^2 \right)$ $\equiv$ $\left( 3J_z^2 - \bm{J}^2, \sqrt{3}(J_x^2 -J_y^2) \right)/2$. 

Very recently, details on the overview of the low temperature properties in PrIr$_2$Zn$_{20}$ have been reported\cite{Onimaru2016PRB, Onimaru2016JPSJ}. The report shows that the NFL behaviors of electrical resistivity and specific heat are very robust against magnetic fields, and their temperature dependences can be well scaled by a characteristic temperature. The satisfaction of the scaling law means that the NFL behavior is governed by the physics with a single energy scale. Interestingly, these temperature dependences are reproducible with Eqs.~\eqref{eq:tsuruta} and \eqref{eq:tsuruta_C}. From these results together with the fact that the quadrupole moment is predominantly active in PrIr$_2$Zn$_{20}$, it is concluded that the NFL behavior in this compound would originate from the quadrupole Kondo effect~\cite{comment2}. Moreover, the report also shows that an exotic FL state with suggestive mass enhancement emerges in the vicinity of the critical point where the quadrupolar ordered phase is fully suppressed by a magnetic field~\cite{Onimaru2016PRB}. This FL state probably serves as an electronic ground state in the quadrupole Kondo lattice to release the residual entropy. However, the formation of the FL state from the NFL state with the strong dumping of quasi-particles is very strange. Some kind of origin to revive the quasi-particles would exist in PrIr$_2$Zn$_{20}$.

Unfortunately, it is still a question whether these findings in PrIr$_2$Zn$_{20}$ are certainly derived from the non-Kramers doublet. To examine this issue, it is necessary to investigate the low temperature properties of other systems under magnetic fields. Thus, we focus on PrRh$_2$Zn$_{20}$, which is a related compound of PrIr$_2$Zn$_{20}$. Although most of the Pr 1-2-20 systems have the well-defined four CEF levels $\Gamma_{1g}(1) \oplus \Gamma_{3g}(2) \oplus \Gamma_{4u}(3) \oplus \Gamma_{5g}(3)$ arising from the local point symmetry $T_d$ at the Pr$^{3+}$ site and the non-Kramers $\Gamma_3$ doublet is located at the ground state, PrRh$_2$Zn$_{20}$ has a different type of the four CEF levels $\Gamma_{1g}(1) \oplus \Gamma_{23g}(2) \oplus \Gamma_{4u}(3) \oplus \Gamma_{4g}(3)$ at low temperatures. This is because the local symmetry of the Pr$^{3+}$ site in PrRh$_2$Zn$_{20}$ is reduced from $T_d$ to $T$ owing to the structural transition around room temperature driven by a low-energy anharmonic vibration of the Zn ions~\cite{Onimaru2012PRB, Wakiya2016PRB}. The CEF level scheme of PrRh$_2$Zn$_{20}$ is determined by the inelastic neutron scattering (INS) experiments as $\Gamma_{23g}$(0~K) - $\Gamma_{4u}$(31.0~K) - $\Gamma_{4g}$(67.1~K) - $\Gamma_{1g}$(78.5~K) in ascending order~\cite{Iwasa2013JPSJ}. Thus, even in this case, PrRh$_2$Zn$_{20}$ can be regarded as a non-Kramers system because the wave function of $\Gamma_{23g}$ is actually equivalent to that of $\Gamma_{3g}$. It has been reported that PrRh$_2$Zn$_{20}$ exhibits a quadrupolar ordering with a large magnetic anisotropy at $T_\mathrm{Q}\sim 0.06$~K~\cite{Onimaru2012PRB}. The ordered phase is deduced to be antiferro-quadrupolar (AFQ) from the negative quadrupole-quadrupole coupling constant~\cite{Ishii2013PRB}. Interestingly, the superconductivity occurs simultaneously with the AFQ order within the experimental accuracy. The superconductivity is so fragile that it collapses under weak magnetic fields of several mT.

There have been only a few intensive studies on PrRh$_2$Zn$_{20}$ reported so far. Compared with the other Pr 1-2-20 systems, most of the low temperature properties of PrRh$_2$Zn$_{20}$ have not been clarified. In this work, we have two objectives: one is to clarify the low temperature properties of PrRh$_2$Zn$_{20}$ under a magnetic field from the viewpoint of transport coefficients such as the electrical resistivity $\rho$ and the Seebeck coefficient $S$. The other is to discuss the expected features throughout non-Kramers systems by comparing the resultant properties of PrRh$_2$Zn$_{20}$ with the report on PrIr$_2$Zn$_{20}$.

In this paper, we begin with the experimental procedure in Sect.~\ref{sec:experiment}. The obtained results of PrRh$_2$Zn$_{20}$ are discussed in detail in Sect.~\ref{sec:result}. In Sect.~\ref{sec:comparison}, we compare the low temperature properties of PrRh$_2$Zn$_{20}$ and PrIr$_2$Zn$_{20}$. In the final section, we summarize this paper.

\section{Experiments}
\label{sec:experiment}
\begin{table}[!t]
\caption{Sizes and residual resistivity ratios of the samples used in this study and direction of applied magnetic field.}
\label{tab:samples}
\begin{tabular}{cccc}\hline\hline
Sample & Size (H$\times$W$\times$L mm$^3$) & RRR & Field direction\\\hline
\#15\_1\_AN & 0.08$\times$0.19$\times$0.47 & 300 & $\bm{B}\parallel [100]$\\
\#15\_2 & 0.05$\times$0.095$\times$0.16 & 15 & $\bm{B}\parallel [100]$\\
\#18 & 0.13$\times$0.26$\times$3.67 & 270 & $\bm{B}\parallel [110]$\\\hline\hline
\end{tabular}
\end{table}
We prepared three types of single-crystalline sample of PrRh$_2$Zn$_{20}$ as shown in Table~\ref{tab:samples}. They were grown by the melt-growth method. The samples named \#15\_1\_AN and \#15\_2 were cut from the same batch, but only \#15\_1\_AN was annealed at 300 $^\circ$C for 4 days in vacuum to improve the crystalline quality. Indeed, the residual resistivity ratio (RRR), which was estimated from the quotient of the resistivity at room temperature and the extrapolated value at 0~K, became markedly higher than that of the unannealed sample \#15\_2. The other sample named \#18 was cut from a different batch from the pair of \#15. The quality of the pristine sample \#18 deduced from the RRR was originally very good and comparable to that of the annealed sample \#15\_1\_AN. The lengthwise directions of samples \#15 and \#18 were oriented in the [100] and [110] directions, respectively. Therefore, samples \#15\_1\_AN and \#15\_2 allow us to investigate the impurity effect, and samples \#15\_1\_AN and \#18 allow us to investigate the anisotropy of the low temperature properties of PrRh$_2$Zn$_{20}$ with respect to the field direction.

The electrical resistivity $\rho$ was measured by a standard four-probe method. The Seebeck coefficient $S$ was measured by a steady-state method. We prepared contacts by spot welding in order to reduce the contact resistance. The samples were fixed on an oxygen-free copper cell attached to the mixing chamber of a $^3$He-$^4$He dilution refrigerator. We carried out the experiments in the temperature range of 0.04--4~K, and the magnetic field range of 0--9~T for samples \#15\_1\_AN and \#15\_2 and the magnetic field range of 0--16~T for sample \#18 by using a 16~T superconducting magnet. 

\section{Results and Discussion}
\label{sec:result}
\subsection{Phase diagram}
\label{subsec:phase diagram}
\begin{figure}[!b]
\centering
\includegraphics[width=240pt]{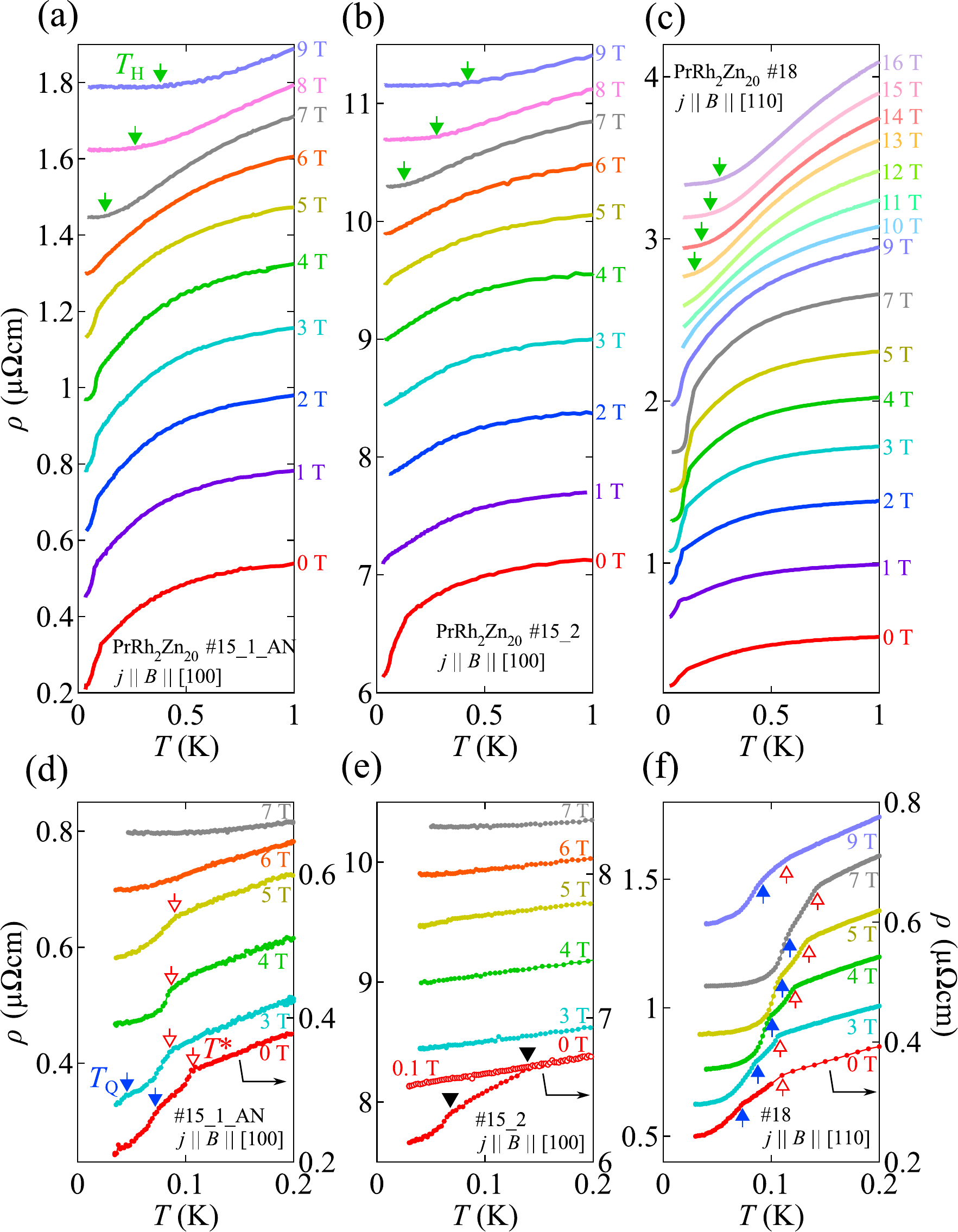}
\caption{(Color online) (a)--(c) Temperature dependences of resistivity of samples \#15\_1\_AN, \#15\_2, and \#18 below 1~K and up to 9 or 16~T. For clarity, the data of (a) and (c) are vertically shifted upward by 0.15~$\mu \Omega$cm relative to each other with increasing field. The arrows indicate the crossover points $T_\mathrm{H}$ between the NFL and FIS states. (d)-(f) Low temperature part of resistivity below 0.2~K under a magnetic field below 0.1~T (right axis) and under a magnetic field above 3~T (left axis). The data of (d) and (f) above 3~T are vertically shifted upward by 0.1~$\mu \Omega$cm relative to each other with increasing field. The solid and open arrows point at the critical temperatures of the AFQ ordered state $T_\mathrm{Q}$ and the HF state $T^*$, respectively. Note that the origin of the anomalies indicated with solid triangles for sample \#15\_2 is unclear at present.}
\label{fig:rhovsT_high}
\end{figure}
First, we construct the temperature-field phase diagram of PrRh$_2$Zn$_{20}$ for all samples from the results of resistivity measurements. Figures~\ref{fig:rhovsT_high}(a)--\ref{fig:rhovsT_high}(c) show the temperature dependences of resistivity under magnetic fields up to 9 or 16~T below 1~K. One can easily find that all of the samples exhibit the NFL behavior with a convex temperature dependence. On the other hand, the NFL behavior gradually changes into a FL-like behavior showing a very weak temperature dependence with increasing field. This is because the non-Kramers doublet is quadratically split by the magnetic field via coupling with the first excited magnetic triplet. Thus, the field-induced singlet (FIS) state is stabilized in the high-field region. The FIS state appears below the crossover temperature $T_\mathrm{H}$, which is defined as the temperature where the resistivity becomes almost flat against temperature, as shown in Figs.~\ref{fig:rhovsT_high}(a)--\ref{fig:rhovsT_high}(c). The observable NFL behavior at lower fields indicates that the splitting is insufficient to form the FIS state in such lower fields. Note that the NFL behaviors on samples \#15\_1\_AN and \#15\_2  gradually change into the FL behavior above 6~T, whereas the NFL behavior on sample \#18 changes into the FL behavior above 10~T. This result evidently demonstrates that the magnetic response of the NFL behavior is anisotropic with respect to the field orientation. The origin of this anisotropy will be discussed in Sect.~\ref{subsec:NFL}. 

The low temperature part of the resistivity shows that there are two kinds of anomalies observed at $T_\mathrm{Q}$ and $T^*$, as shown in Figs.~\ref{fig:rhovsT_high}(d)--\ref{fig:rhovsT_high}(f). Note that the zero resistivity associated with superconductivity is not observed because the superconductivity is already broken by the residual field at several mT of the superconducting magnet. The anomaly at $T_\mathrm{Q}$ is observed in zero field at $\sim$ 0.07~K, which is comparable to the AFQ phase transition temperature reported in Ref.~\citen{Onimaru2012PRB}. Another anomaly at $T^*$ is observed at $\sim 0.1$~K at $B=0$. The anomaly at $T^*$ is interrupted at the finite temperature in the high-field region as shown Fig.~\ref{fig:rhovsT_high}(d), while an AFQ phase boundary is not generally terminated at finite temperature and magnetic field. This indicates that the anomalies at $T^*$ are not associated with the AFQ phase transition. Moreover, the formation of a novel heavy-fermion (HF) state below $T^*$, which is different from an AFQ phase, has been proposed as reported in Ref.~\citen{Yoshida2015PP}. The detailed nature of this anomaly at $T^*$ will be discussed in Sect.~\ref{subsec:GS}. Comparison of samples \#15\_1\_AN and \#15\_2 reveals that the former, i.e., the high-quality sample, shows the anomalies at $T_\mathrm{Q}$ and $T^*$, while the latter, i.e., the low-quality sample, does not show both anomalies, as shown in Figs.~\ref{fig:rhovsT_high}(d) and \ref{fig:rhovsT_high}(e). Thus, the anomalies would be smeared out by lattice disorders in the crystal. Since the lattice disorder causes the non-Kramers doublet to split locally, the long-range order such as the AFQ phase is affected. Therefore, the anomaly at $T_\mathrm{Q}$ in sample \#15\_2 would not be observable. Considering the fact that the anomaly at $T^*$ vanishes as well as that at $T_\mathrm{Q}$, the HF state is also sensitive to lattice disorders. However, the reason why the anomalies of sample \#15\_2 are observable only at $B=0$ is unclear at present. There is a possibility that accidentally contained impurities undergo a superconducting transition~\cite{Onimarupc}; thus, we must be careful of the origin of the anomalies in sample \#15\_2. On the other hand, the NFL behavior is less affected by lattice disorders as shown in Figs.~\ref{fig:rhovsT_high}(a) and \ref{fig:rhovsT_high}(b), suggesting that the local splitting of the non-Kramers doublet in sample \#15\_2 is sufficiently small to be ignored at $T\gtrsim 0.2$~K, where the NFL behavior is observed. In addition, sample \#18 shows the clear double anomalies at $T_\mathrm{Q}$ and $T^*$ up to 9~T. Figures~\ref{fig:rhovsT_high}(d) and \ref{fig:rhovsT_high}(f) clearly show a large anisotropy of $T_\mathrm{Q}$ and $T^*$ with respect to the field direction. In fact, the large anisotropy of the AFQ phase has already been reported~\cite{Onimaru2012PRB, Ishii2013PRB}.

\begin{figure}[!b]
\centering
\includegraphics[width=240pt]{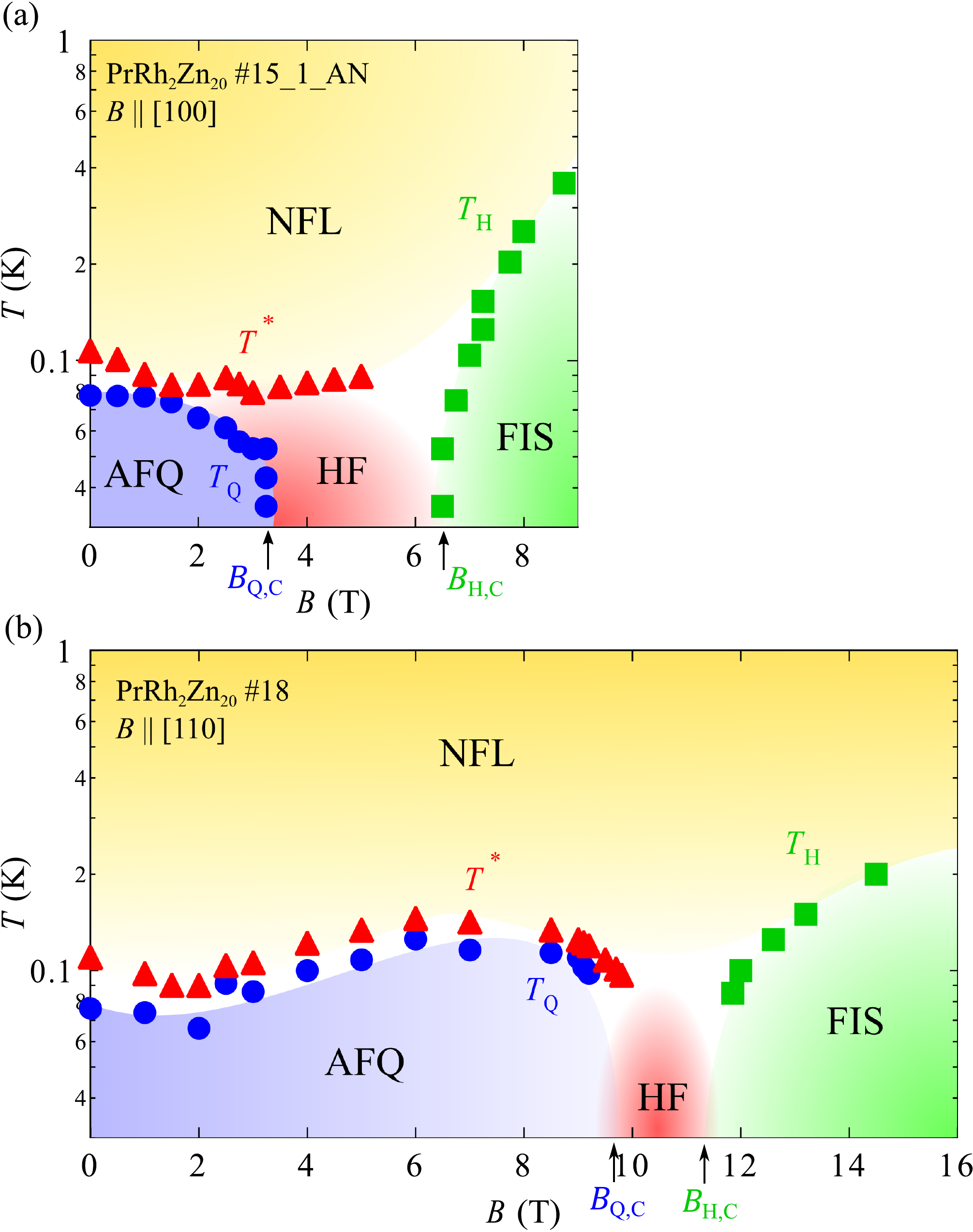}
\caption{(Color online) Temperature-field phase diagrams of PrRh$_2$Zn$_{20}$ for (a)$\bm{B}\parallel [100]$ and (b)$\bm{B}\parallel[110]$. $T_\mathrm{Q}$, $T^*$, and $T_\mathrm{H}$ represent the boundaries or characteristic temperatures of AFQ, HF, and FIS states, respectively. $B_\mathrm{Q,C}$ and $B_\mathrm{H,C}$ are critical fields of the AFQ and FIS states, respectively. The critical fields are estimated to be $B_\mathrm{Q,C} \sim 3.5$~T and $B_\mathrm{H,C} \sim 6.7$~T for $\bm{B}\parallel [100]$, and $B_\mathrm{Q,C} \sim 9.7\pm 0.4$~T and $B_\mathrm{H,C} \sim 11.4 \pm 0.4$~T for $\bm{B}\parallel [110]$.}
\label{fig:phasediagram}
\end{figure}
These features mentioned above are summarized in Fig.~\ref{fig:phasediagram} as a temperature-field phase diagram. Regardless of the field direction, the low temperature electronic states of PrRh$_2$Zn$_{20}$ can be categorized into four states: the NFL state, the AFQ ordered state, the HF state, and the FIS state. $B_\mathrm{Q,C}$ and $B_\mathrm{H,C}$ represent the deduced critical fields of the AFQ phase and FIS state at $T=0$, respectively. We do not illustrate the phase diagram of sample \#15\_2 here because the lattice disorder smears the anomalies associated with the AFQ phase and HF state as mentioned above. The phase diagram is clearly different from the Doniach phase diagram that describes many strong correlated systems with spin degrees of freedom, but bears a close resemblance to the phase diagram of PrIr$_2$Zn$_{20}$~\cite{Yoshida2015PP, Onimaru2016PRB}. This fact implies that the low temperature properties of non-Kramers systems are described in terms of not a mere analogy of the Doniach picture for the Kramers system but another general framework characteristic of the non-Kramers doublet. The detailed property of these states will be discussed in the following sections. 

\subsection{Non-Fermi liquid behavior}
\label{subsec:NFL}
We shall focus on the NFL behavior of PrRh$_2$Zn$_{20}$ in this section. As shown in Fig.~\ref{fig:phasediagram}, the NFL behavior can be observed over a very wide magnetic field range in the high-temperature region above the low temperature electronic states such as the AFQ phase. The exponent of the temperature dependence of $\rho$ shown in Figs.~\ref{fig:rhovsT_high}(a)--\ref{fig:rhovsT_high}(c) is clearly less than unity, implying that the formation of quasi-particles is prevented as mentioned in Sect.~\ref{sec:intro}. Although it has been pointed out that an anharmonic rattling motion characteristic of  caged compounds causes the similar temperature dependence of resistivity~\cite{Hiroi2005JPSJ, Dahm2007PRL}, we can exclude this possibility because the phonon excitation has not been reported below 1~K in some Pr 1-2-20 compounds~\cite{Wakiya2015JPCS, Wakiya2016PRB}. Moreover, the other non-Kramers compound PrPb$_3$ without a caged structure also shows the similar temperature dependence of $\rho$~\cite{Yoshida2016JPCS}. These facts clearly indicate that the convex temperature dependence does not result from the anharmonic phonon due to the characteristic caged structure. Additionally, the NFL behavior in PrRh$_2$Zn$_{20}$ should be completely differentiated from that in the vicinity of a magnetic quantum critical point (QCP). The NFL behavior driven by the spin fluctuations is usually observed within only a narrow magnetic field range around the QCP, whereas our results show the strong robustness of the NFL behavior against a magnetic field. 
\begin{figure}[!b]
\centering
\includegraphics[width=240pt]{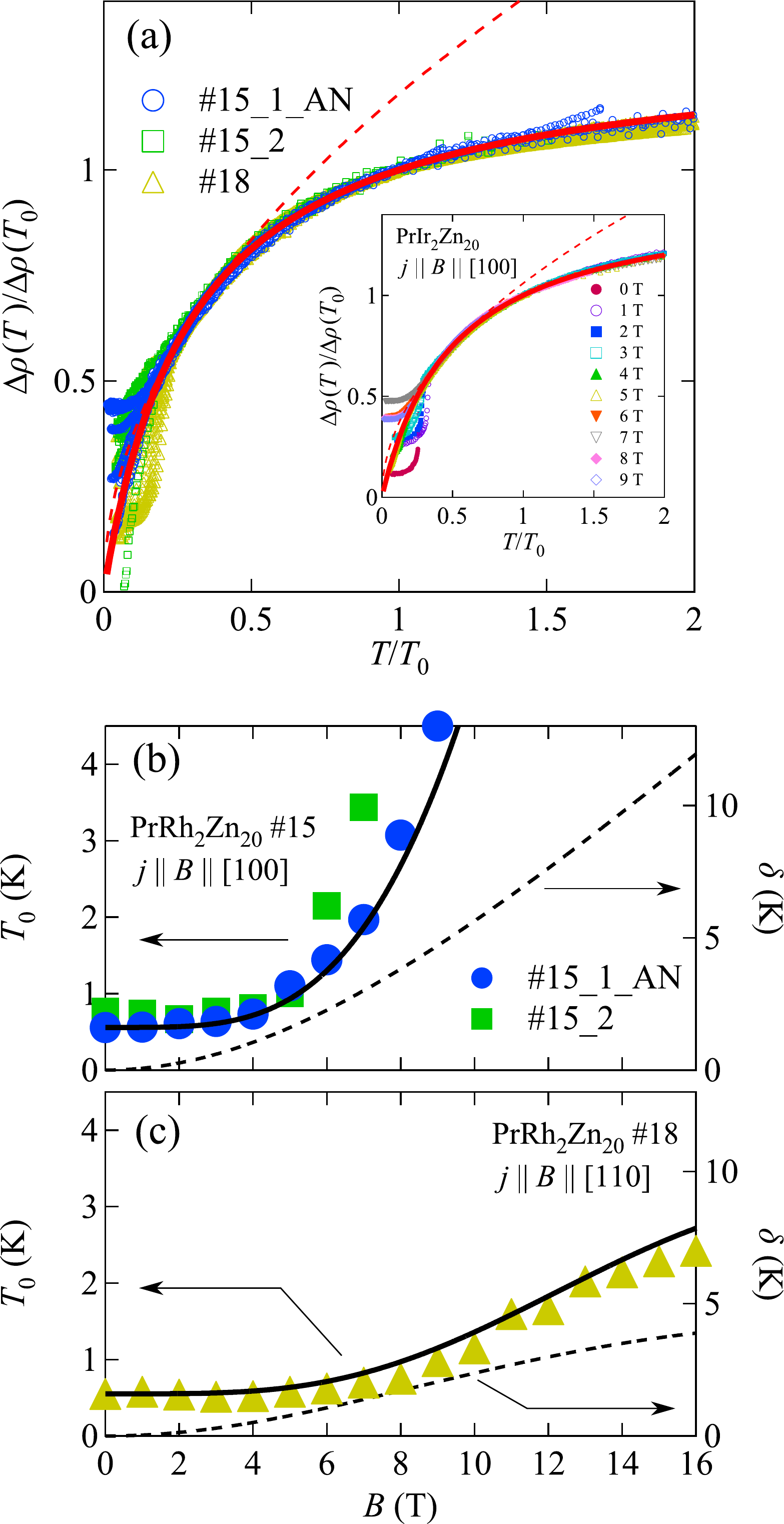}
\caption{(Color online) (a) Universal scaling relation of resistivity in the NFL region. The plotted data includes all of the temperature dependences shown in Figs.~\ref{fig:rhovsT_high}(a)--\ref{fig:rhovsT_high}(c). The open circle, square, and triangle represent samples \#15\_1\_AN, \#15\_2, and \#18, respectively. The thick solid line indicates the plot of Eq.~\eqref{eq:tsuruta} with $a_2=0.3$, and the thin dashed line indicates the plot of $\Delta \rho \propto \sqrt{T}$. The inset shows the scaling relation for PrIr$_2$Zn$_{20}$ with the plots of Eq.~\eqref{eq:tsuruta} with $a_2=0.5$ (solid line) and $\Delta \rho \propto \sqrt{T}$ (dashed line). By using $T_0$ at $B=0$, which is determined in accordance with Cox's definition~\cite{Onimaru2016JPSJ}, $T_0$ under a magnetic field is estimated by fitting Eq.~\eqref{eq:tsuruta}. (b)(c) Field dependences of the characteristic temperature $T_0$. The filled circle, square, and triangle indicate the $T_0$ of samples \#15\_1\_AN, \#15\_2, and \#18, respectively. The solid line (left axis) is the plot of the empirical relation Eq.~\eqref{eq:TK} with the common value $\alpha=0.26$~K$^{-2}$, and the dashed line (right axis) is the splitting of the non-Kramers doublet $\delta (B)$ calculated from the Hamiltonian \eqref{eq:hamiltonian}.}
\label{fig:TKscaling}
\end{figure}

Comparison of Figs.~\ref{fig:rhovsT_high}(a) and \ref{fig:rhovsT_high}(b) reveals that the temperature dependences of resistivity are apparently similar except for the presence or absence of the anomalies at low temperatures. This suggests that the NFL behavior is less sensitive to the lattice disorder leading to a local splitting of the non-Kramers doublet. On the other hand, as seen in Figs.~\ref{fig:rhovsT_high}(a) and \ref{fig:rhovsT_high}(c), the convex temperature dependences for both $\bm{B}\parallel [100]$ and $\bm{B}\parallel [110]$ are shifted to the high-temperature region with increasing field, but the field responses of the shift are quantitatively different from each other. This is nothing but the NFL behavior being governed by an energy scale having an anisotropy with respect to the field direction.

To understand the nature of the NFL behavior, we examine the quadrupole Kondo lattice model and Eq.~\eqref{eq:tsuruta}. Note that we can estimate only $a_2 T_0$ by fitting Eq.~\eqref{eq:tsuruta} to the resistivity, whose value is approximately $a_2 T_0 \sim 0.17$~K at $B=0$. We determine the constant $a_2$ to be $\sim 0.3$ because $T_0$ is estimated in accordance with the definition proposed by Cox, namely, $S_{4f}(T_0)=0.75 R\log 2$, to be $T_0 \sim 0.55$~K at $B=0$~\cite{Cox1998AP, Onimaru2012PRB}. Here, note that the quadrupole Kondo lattice model reproduces the resistivity data in a wider temperature range than the impurity two-channel Kondo model, $\rho-\rho_0 \propto \sqrt{T}$, as shown by the thick solid and thin dashed lines in Fig.~\ref{fig:TKscaling}(a), respectively. By using this value $a_2 \sim 0.3$, we estimate $T_0$ by fitting Eq.~\eqref{eq:tsuruta} under a magnetic field. Interestingly, Fig.~\ref{fig:TKscaling}(a) shows that the resistivities for samples \#15\_1\_AN, \#15\_2, and \#18 in various magnetic fields satisfy the scaling relation in terms of $T_0$. Thus, the scaling relation is widely valid regardless of sample quality and field direction. Moreover, the fact that the scaling also works for PrIr$_2$Zn$_{20}$ as shown in the inset of Fig.~\ref{fig:TKscaling}(a) leads us to state that the NFL behavior observed in the Pr 1-2-20 system is governed by unique physics with the single energy scale $T_0$.

Let us now look at $T_0$ estimated from the scaling in detail. As shown in Figs.~\ref{fig:TKscaling}(b) and \ref{fig:TKscaling}(c), the $T_0$ values of all the samples are increased monotonically with increasing field. In particular, the $T_0$ values of samples \#15\_1\_AN and \#15\_2 vary similarly as shown in Fig.~\ref{fig:TKscaling}(b), indicating that $T_0$ is insensitive to the lattice disorder. On the other hand, as shown in Fig.~\ref{fig:TKscaling}(c), the $T_0$ value of sample \#18 varies weakly in comparison with the other samples. This fact indicates that the anisotropy of the NFL behavior originates from that of $T_0$. One may expect that the upward and anisotropic field dependences of $T_0$ are derived from the field variation of the $c$-$f$ exchange interaction $J_\mathrm{ex}$ because $T_0$ basically corresponds to the Kondo temperature for the two-channel Anderson lattice model; $T_0 \sim T_\mathrm{K} \propto \exp \left[ -1/(N_0 J_\mathrm{ex}) \right]$. However, applying a magnetic field immediately lifts the degeneracy of the non-Kramers doublet and consequently weakens the channel equivalency of the scattering of the conduction electrons. Thus, we should consider the effect of the splitting of the non-Kramers doublet, instead of the field response of $J_\mathrm{ex}$, to clarify the field variation of $T_0$. Here, we assume the following relation;
\begin{equation}
T_0 (B)=T_K \left[ 1 + \alpha \left\{\delta (B)\right\}^2 \right],
\label{eq:TK}
\end{equation}
where $\alpha$ is an arbitrary constant independent of magnetic field. $\delta (B)$ is the splitting of the non-Kramers doublet by a magnetic field. This expression describes that the characteristic temperature $T_0$ agrees with the Kondo temperature $T_K$ at zero field and it deviates from $T_K$ at finite fields in accordance with the splitting $\delta$, which can be calculated by diagonalizing the following Hamiltonian,
\begin{align}
\mathscr{H}=& W\left[\frac{x}{60}\left( O_4^0 + 5 O_4^4 \right)+\frac{1-|x|}{1260}\left( O_6^0 -21 O_6^4 \right)\right.\notag\\
&\quad\left.+\frac{y}{30}\left( O_6^2 - O_6^6 \right) \right]-g_J \mu_B \bm{J}\cdot \bm{B}.
\label{eq:hamiltonian}
\end{align}
The first term is the CEF Hamiltonian for the point group $T$~\cite{Takegahara2001JPSJ}, and the second term is the Zeeman term. We chose the experimentally determined CEF parameters $(W, x, y)=(-1.06, 0.417, 0.0575)$, where $W$ is in units of Kelvin~\cite{Iwasa2013JPSJ}. By using Eq.~\eqref{eq:TK}, we succeeded in reproducing the overall feature of the field dependence of $T_0$ for $\bm{B}\parallel [100]$ and $\bm{B}\parallel [110]$, as shown in Figs.~\ref{fig:TKscaling}(b) and \ref{fig:TKscaling}(c), respectively. It should be emphasized here that the field variation of $T_0$ is reproducible with a common value of the constant $\alpha =0.26$~K$^{-2}$ regardless of the field direction, indicating that the field response of $T_0$ is strictly governed by only the splitting of the non-Kramers doublet $\delta$. In other words, our analysis reveals that the NFL behavior is definitely derived from the non-Kramers doublet. Furthermore, recalling the fact that Eq.~\eqref{eq:tsuruta} is based on the two-channel Anderson lattice model as mentioned in Sect.~\ref{sec:intro}, it can be said that the promising origin of the NFL behavior is the two-channel Kondo effect, namely, the quadrupole Kondo effect.

\subsection{Possible ground states}
\label{subsec:GS}
To release the residual entropy in the quadrupole Kondo lattice as mentioned in Sect.~\ref{sec:intro}, the electronic state must change from the NFL state into another state with decreasing temperature. One of the well-known ground states in non-Kramers systems is a quadrupolar ordered state. In fact, we observed the anomalies of the resistivity associated with the AFQ phase transition at $T_\mathrm{Q}$, and we found that the transition is clearly anisotropic, as shown in Figs.~\ref{fig:rhovsT_high}(d)--\ref{fig:rhovsT_high}(f). With increasing magnetic field, the transition temperature $T_\mathrm{Q}$ for $\bm{B} \parallel [100]$ monotonically decreases to zero at $B_\mathrm{Q,C}^{[100]} \sim 3.5$~T, while $T_\mathrm{Q}$ for $\bm{B} \parallel [110]$ shows a reentrant behavior before decreasing to zero at $B_\mathrm{Q,C}^{[110]} \sim 9.7 \pm 0.4$~T, which is larger than $B_\mathrm{Q,C}^{[100]}$, as shown in Fig.~\ref{fig:phasediagram}. The reentrant structure of  the AFQ phase is usually explained by the fact that the coupling of the field-induced magnetic moment stabilizes the AFQ ordering~\cite{Shiina1997JPSJ, Tayama2001JPSJ}. However, $T_\mathrm{Q}$ for $\bm{B} \parallel [100]$ does not show the reentrant behavior even though the magnetic moment is expected to be induced by applying a magnetic field in the [100] direction~\cite{Hattori2014JPSJ}, suggesting difficulties in clarifying the AFQ phase by simple descriptions. In fact, it has been reported that a thermal fluctuation drives the reentrant behavior of the AFQ phase without any interactions of field-induced moments~\cite{Hattori2016JPSJ}. On the other hand, the anisotropy of $B_\mathrm{Q,C}$ can be explained by that of the energy splitting of the non-Kramers doublet for each field direction. The magnitude of the splitting $\delta$ is estimated by the second-order perturbation to be $\delta^{[100]}>\delta^{[110]}$. Considering that the ordered phase collapses if the energy splitting $\delta$ exceeds the interaction between localized moments, the anisotropy of the AFQ phase is qualitatively explained by that of the splitting $\delta$. 

\begin{figure}[!b]
\centering
\includegraphics[width=240pt]{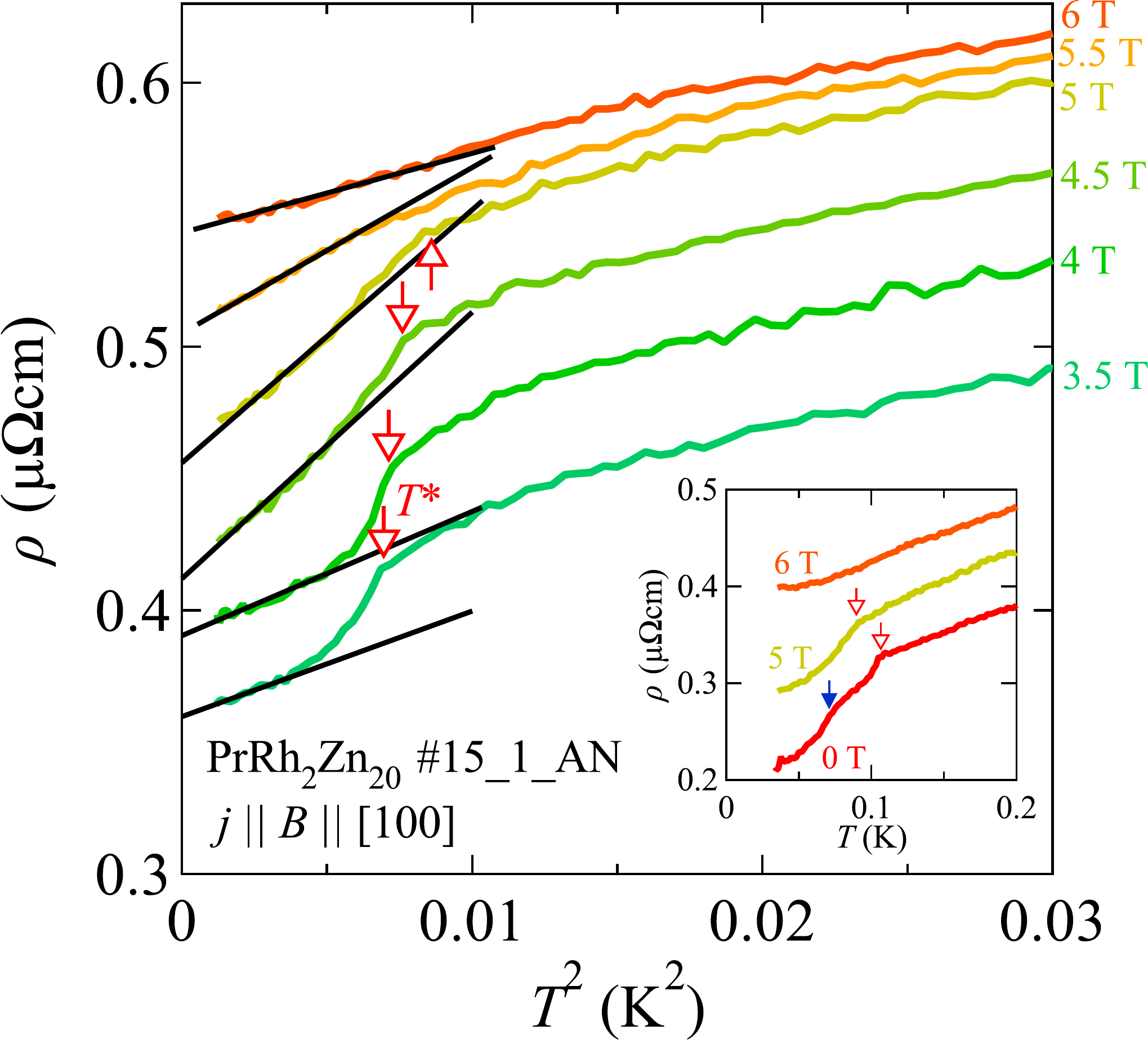}
\caption{(Color online) Resistivity in the HF state of sample \#15\_1\_AN as a function of $T^2$. The open arrows point at the anomaly associated with the HF state, and the straight lines represent the linear fitting at the low temperature limits. The data are vertically shifted upward by 0.03~$\mu \Omega$cm relative to each other with increasing field for clarity. The inset shows $\rho$ vs $T$ plot up to 0.2~K at 0, 5, and 6~T. The data at 5~T is vertically shifted downward by 0.09~$\mu\Omega$cm.}
\label{fig:rhovsT2}
\end{figure}
\begin{figure}[!b]
\centering
\includegraphics[width=240pt]{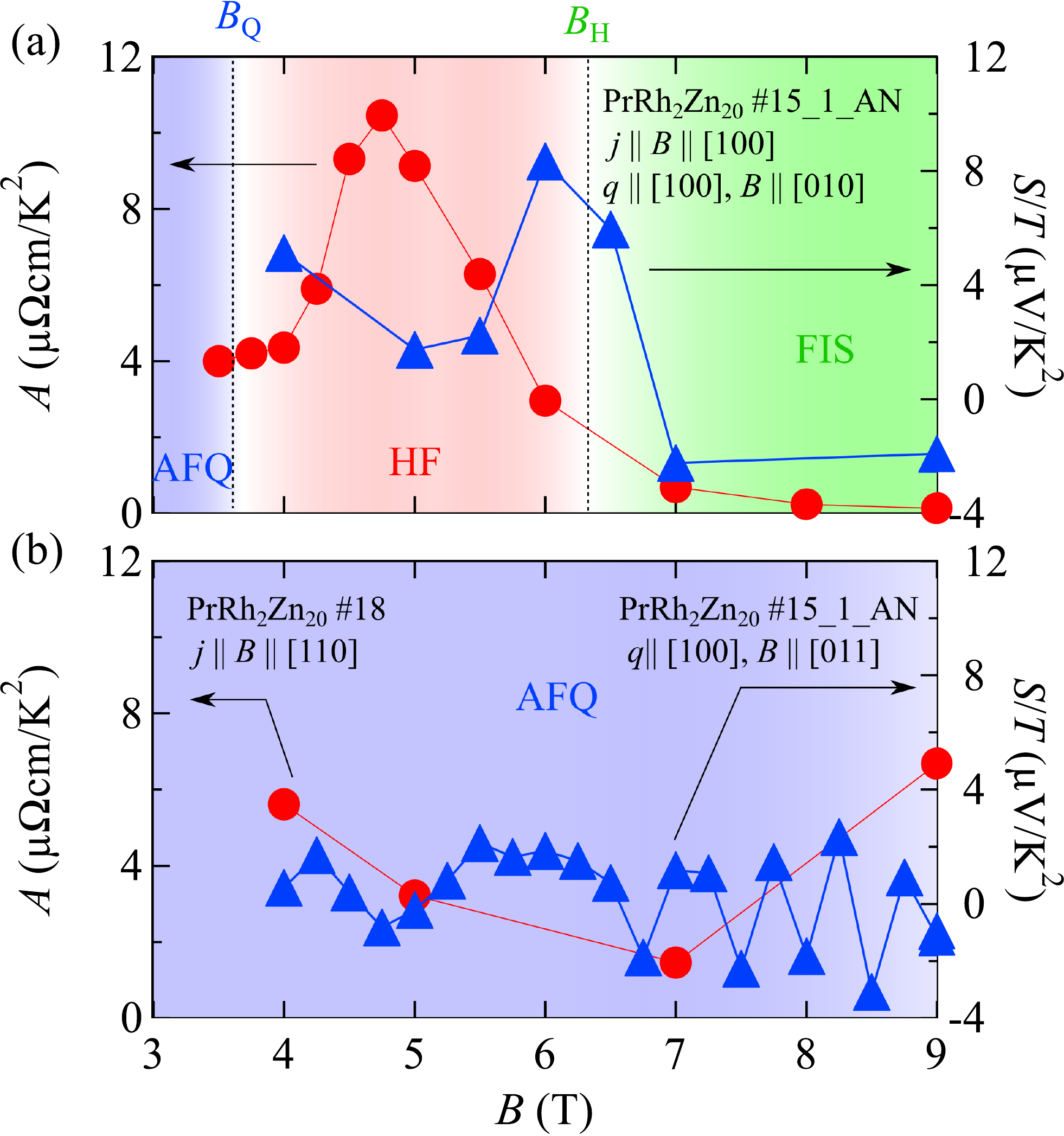}
\caption{(Color online) Field dependence of the $A$ coefficient, which is estimated from the relation $\rho (T)=\rho_0 + A T^2$ (left axis), and the Seebeck coefficient $S$ divided by $T$ at $\sim 70$~mK (right axis) for (a) sample \#15\_1\_AN and (b) sample \#18. Note that $S/T$ for $\bm{B} \parallel [110]$ is measured in sample \#15\_1\_AN. We do not plot the data below 3~T because the $A$ coefficient is difficult to estimate owing to the narrow fitting range~\cite{comment} and the Seebeck coefficient is difficult to measure owing to the small signal.}
\label{fig:AandS_TvsH}
\end{figure}

The other ground state, namely, the HF state, lies in the intermediate field region between $B_\mathrm{Q}$ and $B_\mathrm{H}$ for both field directions. As shown in Fig.~\ref{fig:rhovsT2} and its inset, the resistivity rapidly changes from the NFL behavior into a FL one with decreasing temperature below $T^*$ marked by the open arrows. The solid lines indicate the results of fitting by using the expression $\rho (T)=\rho_0 +A T^2$ at low temperature limits. The emergence of the FL behavior itself is nontrivial and unpredicted within the quadrupole Kondo model. To elucidate this HF state seen below $T^*$, we examined the $A$ coefficient being proportional to ${m^\ast}^2$ ($m^*$: effective mass of electrons) and the Seebeck coefficient. Figure~\ref{fig:AandS_TvsH} shows the magnetic field dependences of the $A$ coefficient and the Seebeck coefficient divided by temperature, $S/T$, for samples \#15\_1\_AN and \#18 at $\sim 70$~mK, ranging from 3 to 9~T. It turns out that the $A$ coefficient of sample \#15\_1\_AN is prominently enhanced at around 4.8~T for $\bm{B} \parallel [100]$, which is between two characteristic fields $B_\mathrm{Q}$ and $B_\mathrm{H}$. On the other hand, the value for sample \#18 is not enhanced at the same field region, where the AFQ phase lies~\cite{comment}. The maximum $A$ value of sample \#15\_1\_AN is approximately $\sim$ 10~$\mu \Omega$cm/K$^2$. This value is comparable to the value of PrIr$_2$Zn$_{20}$ $\sim$ 15~$\mu \Omega$cm/K$^2$, which roughly satisfies the Kadowaki--Woods relation~\cite{Onimaru2016PRB, Machida2015JPCS}. This implies that the nontrivial FL state in PrRh$_2$Zn$_{20}$ also has a strikingly large electron mass. On the other hand, the Seebeck coefficient is relatively small in this field range. Rather, $S/T$ seems to be enhanced on the verge of the crossover field $B_\mathrm{H}$, implying that $S/T$ is enhanced by the marked change in the electronic states around $B_\mathrm{H}$ via the relation $S/T \propto \left. d\sigma/d\varepsilon \right|_{\varepsilon=\varepsilon_F}$. Although $S/T$ is sometimes discussed to be directly related to an electron mass, note that it is not always true.

At a glance, the enhancement of the $A$ coefficient around the critical field $B_\mathrm{Q,C}$ is reminiscent of the quantum critical behavior near a QCP. In fact, the possibility of a quadrupolar quantum critical point (Q-QCP) has been proposed in Pr$Tr_2$Al$_{20}$ systems~\cite{Tsujimoto2014PRL, Matsubayashi2012PRL}. These reports claim that the observations of the superconductivity with sizable mass enhancement in PrTi$_2$Al$_{20}$ and PrV$_2$Al$_{20}$ result from developing the quadrupolar fluctuations on the verge of the FQ (AFQ) ordered phase. However, quantum critical behaviors such as the anomalous critical exponent or critical divergent behavior of physical quantities have not been observed around the critical fields at low temperatures below $T^*$ in Pr$Tr_2$Zn$_{20}$ systems; instead, the FL behavior is observed. It would be unnatural that the FL picture recovers if we consider that the quadrupolar fluctuations are strongly developed in the vicinity of $B_\mathrm{Q,C}$. These considerations lead us to conclude that the Q-QCP scenario is unlikely in Pr$Tr_2$Zn$_{20}$ systems.

We have proposed an orbital selective Kondo effect, that is, a composite order as an origin of this HF state~\cite{Onimaru2016PRB}. According to the theory~\cite{Hoshino2011PRL,  Hoshino2013JPSJ}, the symmetry between equivalent scattering channels of conduction electrons is spontaneously broken as a second-order phase transition so as to release the residual entropy of $\log\!\!\sqrt{2}$ in the two-channel Anderson lattice. In addition, the conduction electron with the decoupled channel shows a FL behavior with enhanced effective mass~\cite{Hoshino2011PRL}. Under a magnetic field, the second-order transition should change into a crossover because the two channels are forced to be inequivalent~\cite{Yotsuhashi2002JPSJ}. If the magnetic field is relatively weak, that is, the channel inequivalence is sufficiently small, it is highly possible that the crossover occurs very rapidly like a phase transition. This scenario is apparently consistent with our results; the resistivity exhibits a FL behavior below the anomaly at $T^*$, and the anomalies at $T^*$ are sharp at low fields but they gradually become ambiguous with increasing field. Moreover, this broadening of the anomaly clearly depends on the field direction, as shown in Figs.~\ref{fig:rhovsT_high}(d) and \ref{fig:rhovsT_high}(f). The $T^*$ for $\bm{B}\parallel [100]$ becomes ambiguous around 6~T, while that for $\bm{B}\parallel [110]$ is observable even at 9~T. Thus, the development of the channel inequivalence expressed by the splitting of the non-Kramers doublet would be essential for the emergence of the HF state. However, we have not obtained direct evidence to realize the composite order below $T^*$.  Further studies, especially microscopic measurements, are strongly required to specify the nature of this novel HF state such as the order parameter.

\begin{figure}[!b]
\centering
\includegraphics[width=240pt]{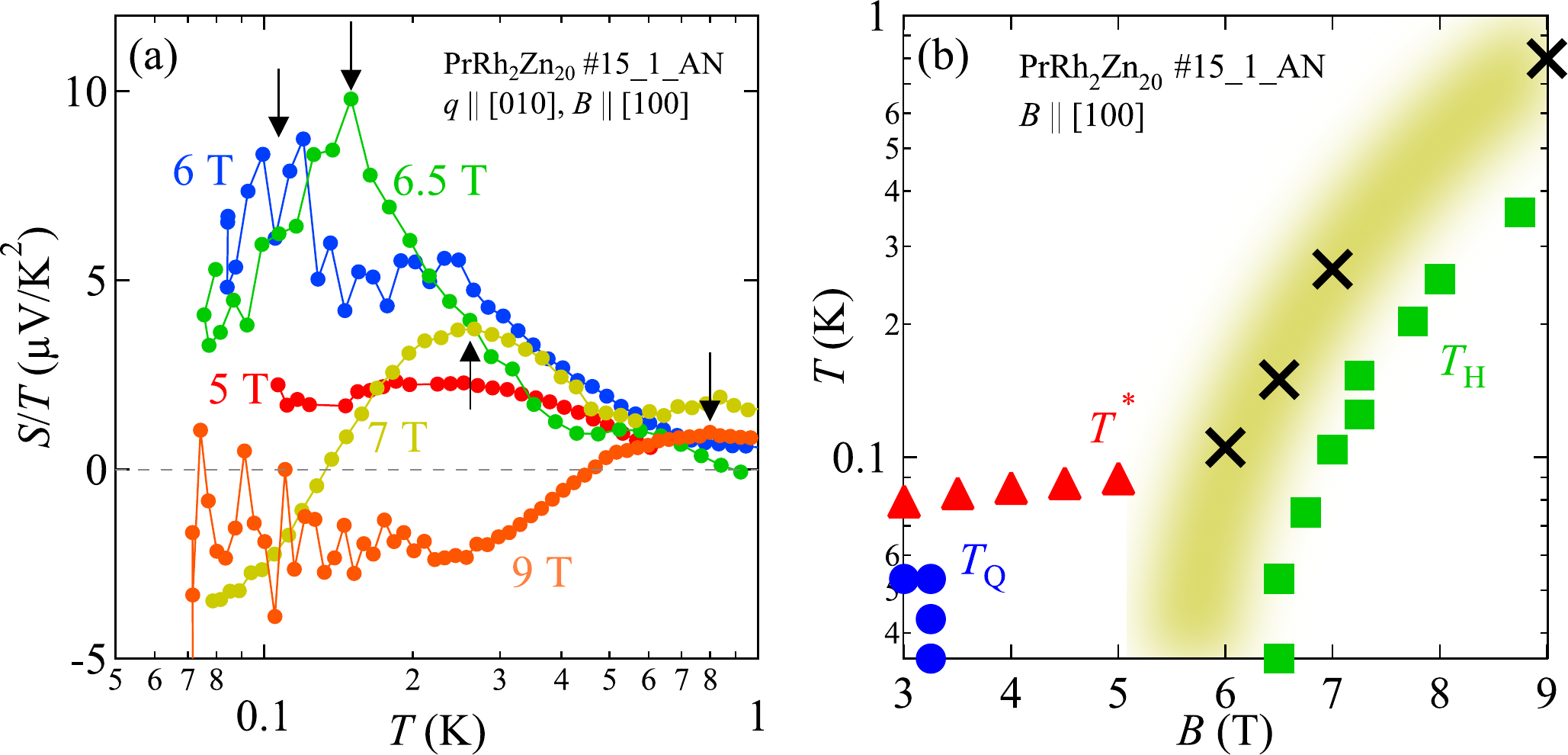}
\caption{(Color online) (a) Temperature dependence of $S/T$ below 1 K at selected fields above 5~T. The black arrows point at the peak position of $S/T$. (b) $B$-$T$ phase diagram of sample \#15\_1\_AN ranging from 3 to 9~T. The cross mark represents the peak position of $S/T$. The shaded region shows an expected distribution of the peak positions of $S/T$.}
\label{fig:S_TvsT}
\end{figure}
By applying a magnetic field above $B_\mathrm{H}$, the ground state changes into the FIS state for both field directions, as shown in Fig.~\ref{fig:phasediagram}. According to the quadrupole Kondo model, the crossover between the NFL and FIS states occurs at $T_\mathrm{H}$ proportional to $\delta^2 \propto B^{2-4}$ at high fields~\cite{Yotsuhashi2002JPSJ, Kusunose2015JPCS}. The $c$-$f$ coupling becomes weak deep inside the FIS state in accordance with the splitting of the non-Kramers doublet, so that the strong electron scattering due to the two-channel Kondo effect is suppressed and the resistivity becomes almost independent of temperature, as shown in Figs.~\ref{fig:rhovsT_high}(a)--\ref{fig:rhovsT_high}(c).

On further applying a magnetic field to sample \#18 in the [110] direction, the NFL behavior begins to be replaced by the FIS state above $B\sim 11$~T, which is larger than $B\sim 7$~T applied in the [100] direction for sample \#15\_1\_AN. The anisotropy of the FIS state would reflect the anisotropic field response of the splitting $\delta$ for $\bm{B}\parallel [100]$ and $\bm{B}\parallel [110]$ because $T_\mathrm{H} \propto \delta^2$ for large $B$. Let us discuss the anisotropy of $B_\mathrm{H,C}$ quantitatively. The field response of $\delta$ can be easily calculated on the basis of the perturbation theory to be $\delta^{[100]} \propto B^2$ and $\delta^{[110]} \propto B^2/2$. If we consider that the splitting $\delta$ takes the same value at $B_\mathrm{H}$ for both field directions, namely, $\delta^{[100]} (B_\mathrm{H,C}^{[100]})= \delta^{[110]} (B_\mathrm{H,C}^{[110]})$, the ratio of the critical fields $B_\mathrm{H,C}$ is simply calculated as
\begin{equation}
\frac{B_\mathrm{H,C}^{[110]}}{B_\mathrm{H,C}^{[100]}} \sim 1.4.
\end{equation}
This estimated value of 1.4 approximately agrees with the ratio of the experimental values  $B_\mathrm{H,C}^{[110]}/B_\mathrm{H,C}^{[100]}=(11.4\pm0.4)/6.7 \sim 1.7\pm0.1$. The values $B_\mathrm{H,C}$ are estimated by the extrapolation of the boundary to $T=0$. A slight discrepancy of the factors might result from the estimation error and/or the limitation of the perturbation theory in this field range.

Although the FIS state itself is simple, the boundary would not be so simple. Recalling that the Seebeck coefficient is markedly enhanced around the boundary of the FIS state $B_\mathrm{H}(T_\mathrm{H})$, as shown in Fig.~\ref{fig:AandS_TvsH}(a), we expect that the electronic state changes abruptly around $B_\mathrm{H}(T_\mathrm{H})$. Actually, there exists the peak structure of $S/T$ along the $B_\mathrm{H}(T_\mathrm{H})$ boundary even at high temperatures and high fields, as shown in Fig.~\ref{fig:S_TvsT}. This supports the fact that the peak position of $S/T$ does not coincide with the position of the $A$ coefficient, as shown in Fig.~\ref{fig:AandS_TvsH}(a). Moreover, the temperature dependence of $S/T$ does not show a divergent tendency at the low temperature limit, as shown in Fig.~\ref{fig:S_TvsT}(a). From these facts, a quadrupolar quantum criticality would unlikely be observed in this system as mentioned earlier. However, we have not revealed the specific mechanism of the peak of $S/T$. Why the Seebeck coefficient is strongly enhanced along $T_\mathrm{H}$ is still an open question to be addressed.

\begin{table*}[t]
\centering
\caption{Crystallographic parameters and characteristic energy scales of PrIr$_2$Zn$_{20}$ and PrRh$_2$Zn$_{20}$. The point group and the energy level scheme are determined by the INS experiments~\cite{Iwasa2013JPSJ}. The lattice constant $a$ and the $\Gamma_3$-type quadrupolar interaction $g'$ are reported in Refs. \citen{Onimaru2010JPSJ, Onimaru2012PRB, Ishii2013PRB}, and \citen{Ishii2011JPSJ}. The listed values of $T_\mathrm{Q}$ and $T^*$ are estimated at $B=0$.}
\label{tab:IrRh}
\begin{tabular}{ccccccccc}\hline\hline
Compound	&Symmetry &Level scheme (K)&$a$ (\AA) & $g'$ (K) & $T_\mathrm{Q}$ (K) & $T^*$ (K) & $B_\mathrm{Q,C}$ (T) & $B_\mathrm{H,C}$ (T) \\\hline
PrIr$_2$Zn$_{20}$ & $T_d$  & $\Gamma_3(0) - \Gamma_4(27.4) - \Gamma_1(65.8) - \Gamma_5(67.3)$ & 14.2729(2) & -0.13 & 0.13 & --- & 4.3 & 5.5 \\
PrRh$_2$Zn$_{20}$& $T$  & $\Gamma_{23}(0) - \Gamma_4^{(1)}(31.0) - \Gamma_4^{(2)}(67.1) - \Gamma_1(78.5)$   &14.2702(3) & -2.328 & 0.07 & 0.1 & 3.5 & 6.7 \\\hline\hline
\end{tabular}
\end{table*}

\section{Comparison with PrIr$_2$Zn$_{20}$}
\label{sec:comparison}
As discussed in the previous section, we found that the low temperature electronic states of PrRh$_2$Zn$_{20}$ are categorized into four regions, and all of these states are characterized by the energy splitting of the non-Kramers doublet. In other words, these four states of PrRh$_2$Zn$_{20}$ are certainly derived from the non-Kramers doublet. It is very important to find the universal features of electronic states in the non-Kramers system to compare our results with those of the other non-Kramers systems. As the first step, let us discuss our data again in comparison with the report on PrIr$_2$Zn$_{20}$~\cite{Onimaru2016PRB, Onimaru2016JPSJ}.

PrIr$_2$Zn$_{20}$, as well as PrRh$_2$Zn$_{20}$, belongs to the Pr 1-2-20 systems as mentioned in Sect.~\ref{sec:compound}. PrIr$_2$Zn$_{20}$ shares similar features to PrRh$_2$Zn$_{20}$, e.g., the CEF first excited energy and the presence of the NFL, AFQ, HF, and FIS states, except for the local point group symmetry. The notable details of both compounds are described in Table~\ref{tab:IrRh}. Hereafter, we will express each quantity such as $T_\mathrm{Q}$ and $B_\mathrm{H,C}$ with a suffix representing the compound, e.g., $T_\mathrm{Q}^\mathrm{Rh}$ and $B_\mathrm{H,C}^\mathrm{Ir}$.

It was found that resistivity commonly exhibits the convex curve NFL behavior for both compounds. The most important is the fact that the NFL behavior can be described by the unique Eq.~\eqref{eq:tsuruta} with a unique characteristic temperature $T_0$ regardless of the compound. Although the characteristic temperatures $T_0$ for the different compounds are naturally different from each other, it must be astounding that the same scaling expression works very well among the different compounds. Moreover, in the other non-Kramers system PrPb$_3$, which has a different cubic structure $Pm\bar{3}m$ from that of Pr 1-2-20 systems, we found that resistivity satisfies the same scaling relation on Eq.~\eqref{eq:tsuruta} as well~\cite{Yoshidapc}. The satisfaction of the scaling for the different structural compounds strongly suggests that non-Kramers systems share a universal mechanism for the NFL behavior, namely, the quadrupole Kondo effect, rather than a detailed mechanism that is valid for only the individual compounds. Although many experiments on the basis of the diluted U- and Pr-based compounds have provided the results hinting at the quadrupole (impurity) Kondo effect~\cite{Seaman1992JAC, Amitsuka1994JPSJ, Kawae2006PRL}, the complexity of the diluted system often prevented the study of the universal behavior, which is a key for the Kondo effect. Therefore, our studies on PrRh$_2$Zn$_{20}$ and PrIr$_2$Zn$_{20}$ provide the systematic results to clarify the essential nature of the quadrupole Kondo effect. For further support of our arguments, we need to perform the same scaling for the other physical quantities such as specific heat and magnetic susceptibility.

Let us now turn our attention to the ground states. The AFQ transition temperature $T_\mathrm{Q}^\mathrm{Rh} \sim 0.07$~K at zero field is approximately half of $T_\mathrm{Q}^\mathrm{Ir} \sim 0.13$~K at zero field, and the critical field $B_\mathrm{Q,C}^\mathrm{Rh}\sim 3.5$~T is also smaller than $B_\mathrm{Q,C}^\mathrm{Ir}\sim 4.3$~T. The difference in the transition temperature $T_\mathrm{Q}$ of the AFQ phase is simply expected to come from that of the RKKY-type quadrupolar interactions for each compound. Since the Ir and Rh ions provide the same number of $d$-electrons, the difference in the quadrupolar interaction would be derived from the chemical compression of the crystal lattice. As shown in Table~\ref{tab:IrRh}, the lattice constant of PrRh$_2$Zn$_{20}$ is slightly smaller than that of PrIr$_2$Zn$_{20}$, so that the effect of the chemical compression on PrRh$_2$Zn$_{20}$ is expected to be stronger. However, the difference in the lattice constant for these compounds is considerably small: $\Delta a/a \sim 0.02$\%, $\Delta V/V \sim 0.06$\%. It is unlikely that this small change in the lattice constant results in the variance of the AFQ phase transition temperature $T_\mathrm{Q}$. Even though we assume that the small change is significant to suppress the $T_\mathrm{Q}$ of PrRh$_2$Zn$_{20}$, it is inconsistent with the result of the recent hydrostatic pressure experiments that $T_\mathrm{Q}$ for both compounds initially increases with increasing pressure~\cite{Umeopc}. Moreover, the ultrasonic experiments have reported that the intersite quadrupolar interaction $g'_\mathrm{Rh}$ is much larger than $g'_\mathrm{Ir}$, although $T_\mathrm{Q}^\mathrm{Rh}$ is smaller than $T_\mathrm{Q}^\mathrm{Ir}$~\cite{Ishii2011JPSJ, Ishii2013PRB}. Therefore, we would not be able to explain the difference in $T_\mathrm{Q}$ in terms of only the quadrupolar interaction. Some kind of suppression mechanism of the AFQ phase may be hidden in PrRh$_2$Zn$_{20}$.

The conspicuous qualitative difference between these compounds is the presence or absence of $T^*$ at zero field. Although both compounds exhibit the FL behavior with a large $A$ coefficient below $T^*$ in a magnetic field, $T^*$ is observable even at $B=0$ only in PrRh$_2$Zn$_{20}$. As discussed in Sect.~\ref{subsec:GS}, we have proposed the composite order as the origin of $T^*$. In this case, the energy scale of the composite order must be superior to that of the AFQ ordering so that $T^*$ is explicitly observed at $B=0$ in PrRh$_2$Zn$_{20}$. According to the theoretical study, however, the composite order is always masked by the quadrupolar ordered phase around the half filling~\cite{Hoshino2013JPSJ}. Hence, some kind of mechanism is required so that the composite order overcomes the AFQ order. One possibility is that the crystal lattice of PrRh$_2$Zn$_{20}$ involves a geometrical frustration or a second nearest-neighbor interaction associated with the point group $T$. These effects naturally suppress the AFQ phase. In fact, it has been theoretically pointed out that the composite order becomes superior to the AFQ phase by taking in these effects~\cite{Hoshino2013JPSJ}. Moreover, if the crystal lattice of PrRh$_2$Zn$_{20}$ involves these effects, the mysterious contradiction between the smaller AFQ transition temperature $T_\mathrm{Q}^\mathrm{Rh}$ and the larger coupling constant $g'_\mathrm{Rh}$ as mentioned in the previous paragraph can be explained qualitatively. Unfortunately, neither the theoretical nor experimental studies supporting such possibilities in PrRh$_2$Zn$_{20}$ have been reported so far. In any case, except for the presence or absence of $T^*$ at $B=0$, it is common that the novel HF state is formed below $T^*$, and $T^*$ gradually becomes ambiguous with increasing field.

Finally, let us discuss the FIS state. As we mentioned in Sect.~\ref{subsec:GS}, a sufficiently high field gives rise to the general FL state due to the large splitting of the non-Kramers doublet. Namely, the FIS state easily emerges if only the non-Kramers doublet is split by even weaker fields. Since the splitting $\delta$ follows $\propto B^2/\Delta$ within the perturbation theory, where $\Delta$ is the first excited energy from the CEF ground state, the compound having the smaller $\Delta$ would form the FIS state at lower fields. In practice, PrIr$_2$Zn$_{20}$ has the smaller $\Delta^\mathrm{Ir} \sim 27.4$~K than the value of PrRh$_2$Zn$_{20}$ $\Delta^\mathrm{Rh} \sim 31.0$~K, and accordingly, PrIr$_2$Zn$_{20}$ forms the FIS state at the lower field $B_\mathrm{H,C}^\mathrm{Ir} \sim 5.5$~T than $B_\mathrm{H,C}^\mathrm{Rh} \sim 6.7$~T, as described in Table~\ref{tab:IrRh}. This is qualitatively consistent with the argument above and in Sect.~\ref{subsec:GS}. Interestingly, in PrIr$_2$Zn$_{20}$, the entry of the FIS state accompanies the remarkable enhancement of the Seebeck coefficient as well as in PrRh$_2$Zn$_{20}$. Very recently, it has been found that the magnetostriction of PrRh$_2$Zn$_{20}$ exhibits a prominent peak along the boundary~\cite{Onimarupc}. This implies, from the viewpoint of both microscopic and macroscopic aspects, that there is a marked change between the quadrupolar many-body state of the NFL or HF state and the simple metallic state of the FIS state. It should be important to pay much attention to the nature of the FIS state and the boundary.

\section{Summary}
\begin{figure}[!b]
\centering
\includegraphics[width=240pt]{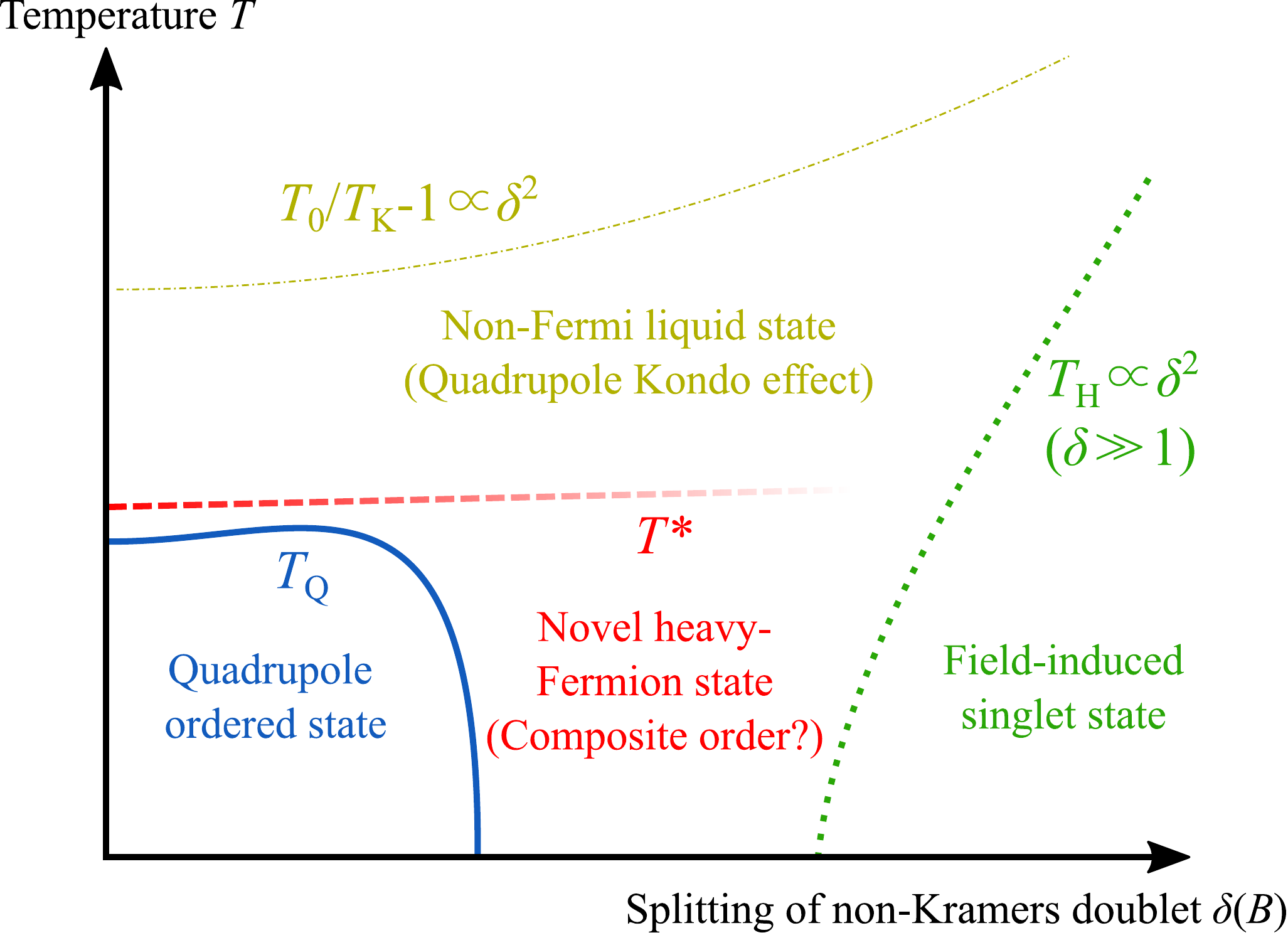}
\caption{(Color online) Schematic phase diagram expected for non-Kramers systems. The solid line $T_\mathrm{Q}$ is the phase boundary of the quadrupole ordered state, and the dashed line $T^*$ and dotted line $T_\mathrm{H}$ are the NFL-HF crossover and NFL-FIS crossover lines, respectively. The thin dash-dotted line $T_0$ is the characteristic temperature of the quadrupole Kondo effect. The horizontal axis is set to be the energy splitting of the non-Kramers doublet by the magnetic field $\delta (B)$, reflecting that these four states are characterized by $\delta (B)$. The existence of $T^*$ at $B=0$ is dependent on the compound.}
\label{fig:genericdiagram}
\end{figure}
We have investigated the quadrupole-driven exotic physics and their magnetic field effect in the non-Kramers system PrRh$_2$Zn$_{20}$ from the viewpoint of transport properties such as electrical resistivity and Seebeck coefficient. As a result, we revealed the anisotropic $B$-$T$ phase diagram composed of the four states for $\bm{B}\parallel [100]$ and $\bm{B}\parallel [110]$: NFL, AFQ, HF, and FIS states.

In the NFL state, the convex temperature dependence of resistivity with the exponent less than unity was observed over a wide field region, reflecting the strong dumping of quasi-particles due to quadrupolar many-body effects. These dependences are well described in terms of the two-channel Kondo lattice model with the characteristic temperature $T_0$. In addition, $T_0$ is well expressed by the empirical relation $T_0(B)=T_K \left[1 +\alpha \left\{ \delta (B) \right\}^2\right]$ with the energy splitting of the non-Kramers doublet $\delta$. This obviously shows that the NFL behavior is strictly governed by the non-Kramers doublet, supporting the quadrupole Kondo lattice scenario. In the AFQ ordered state, the transition temperature $T_\mathrm{Q}$ for $\bm{B}\parallel [100]$ monotonically decreases with increasing field, while $T_\mathrm{Q}$ for $\bm{B}\parallel [110]$ shows a reentrant behavior. The origin of this difference is not clarified yet, although the effects of field-induced moments and thermal fluctuations have been suggested as the candidate for the reentrant feature. In any case, the anisotropy of the critical field $B_\mathrm{Q,C}$ is mainly explained by the difference in the splitting of the non-Kramers doublet. In the HF state below $T^*$, the resistivity exhibits a FL behavior with a large effective mass deduced from the $A$ coefficient of $\Delta \rho (T) = A T^2$, although the quasi-particle picture originally fails above $T^*$. The anomaly between the NFL and HF states becomes broadened with increasing field, and the field-response of the anomaly is anisotropic with respect to the field direction, implying that the channel inequivalence expressed by the splitting of the non-Kramers doublet is crucial for the formation of the HF state. From these facts, we proposed a composite order as the origin of this FL behavior. In the FIS state, the resistivity changes gradually from the NFL behavior to a FL behavior with a very weak temperature dependence at temperatures below $T_\mathrm{H}$. This is because the considerable splitting of non-Kramers doublet causes a singlet state without quadrupole degrees of freedom. Interestingly, it is found that the Seebeck coefficient is strongly enhanced along the crossover temperature $T_\mathrm{H}$, implying that the electronic state abruptly changes around the boundary of the FIS state.

Thus, as discussed in this paper, we revealed that all of the four states constituting the phase diagram of PrRh$_2$Zn$_{20}$ have a close connection with the splitting of the non-Kramers doublet. Considering that a similar phase diagram has been observed in another non-Kramers system PrIr$_2$Zn$_{20}$, the present results strongly suggest that non-Kramers systems share a common phase diagram. In fact, the other non-Kramers system PrV$_2$Al$_{20}$ also shows a similar $B$-$T$ phase diagram~\cite{Shimura2015PRB}. On the basis of these facts, here, we propose a common phase diagram for non-Kramers systems with respect to the energy splitting $\delta (B)$ under a magnetic field, as illustrated in Fig.~\ref{fig:genericdiagram}. Note that the existences of $T^*$ at $B=0$ and the quadrupole ordered structure are dependent on the compound. In addition, the quadrupole ordered state and the novel heavy-fermion state can be easily modified because these states are very sensitive to lattice disorders. To confirm further the certain universality of this proposed phase diagram, further experiments focusing on the other non-Kramers systems, not only Pr 1-2-20 systems, are highly required.

\section*{Acknowledgments}
We would like to thank \mbox{K. Hattori}, \mbox{S. Hoshino}, \mbox{K. Inui}, \mbox{Y. Kuramoto}, \mbox{H. Kusunose}, \mbox{K. Miyake}, \mbox{Y. Motome}, \mbox{A. Tsuruta}, and \mbox{K. Umeo} for valuable discussions. This work was supported by JSPS KAKENHI Grant Numbers 15H05884 (J-Physics) and 16J06992.


\begin{thebibliography}{10}
\bibitem{comment2}
Usually, the Kondo effect is distinctly described by the impurity model or
  lattice model. In this paper, the Kondo effect refers to the one that is
  described by the lattice model, unless otherwise noted.
\bibitem{comment}
We were not able to measure the $A$ coefficient below 3~T precisely because the
  fitting range was strictly limited owing to the very low AFQ phase transition
  temperature $\sim 0.07$~K and the lowest temperature of our experiment $\sim
  0.04$~K.
\end{thebibliography}
\end{document}